**Phase field simulation of grain size effects in nanograined Ti-Nb shape memory alloys**

Jiawei Mai, Yuquan Zhu, Tao Xu*, Tong-Yi Zhang*

Materials Genome Institute, Shanghai University

Shanghai 200444, China

**Abstract**

Titanium-based shape memory alloys, such as Ti2448, have attracted enormous attention owing to their unique thermomechanical properties and potential biomedical applications. In this study, we develop a polycrystalline phase field to investigate the grain size dependence of the martensitic transformation and associated mechanical properties of nanograined Ti-Nb alloys. It is shown that a reduction of the average grain size strengthens the suppression of the martensitic transformation (MT), leading to an increase of the transformation stress, shrinkage of the stress hysteresis, and elimination of residual strain. The time-temperature-transformation curves of nano-grained Ti-Nb alloys with different average grain sizes are obtained and the validity of Hall-Petch relation is also confirmed in all studied grain sizes. Furthermore, when the average grain size becomes ultrasmall, both the temperature- and stress-induced MTs show the continuous second-order phase transition behavior. These superior transformation characteristics are attributed to the high density of grain boundaries and the related dominant role of the gradient energy at the nanoscale. Our results have profound implications for the design and control of the properties in nano-grained shape memory alloys.

---

* Corresponding authors: Tao Xu xutao6313@shu.edu.cn; Tong-Yi Zhang zhangty@shu.edu.cn

## 1. Introduction

Reducing grain size of polycrystalline metals and alloys is the most efficient approach to improve most mechanical properties except of creep resistance, as indicated by the Hall-Petch equation [1-3]. When the grain size of an alloy is smaller than 100 nm, the alloy is called Nanometer-Grained (NG) alloy. Shape memory alloys (SMAs) have found wide applications in various fields due to their unique shape memory effect and pseudoelasticity, which originate from the temperature- or/and stress-induced reversible martensitic transformation (MT) [4]. Extensive efforts [5-12] have endeavored to explore, improve, and tailor the thermal and mechanical properties of SMAs, among which grain refinement [7, 8] exhibits bright prospects.

Experimentally Waitz et al. [13] investigated MT in NG $Ti_{49.7}Ni_{50.3}$ by transmission electron microscopy (TEM). They employed high pressure torsion deformation and following annealing to prepare the NG samples with grain size ranging from 5 to 350 nm. The TEM observations show that upon cooling the nanostructures transform to martensite partially since the grain boundaries hinder the autocatalytic formation of martensite. The energy barrier against MT increases with decreasing grain size and the MT is completely suppressed in the NG SMA with grain size smaller than 60 nm. With differential scanning calorimetry (DSC), TEM, and X-ray diffractometry, Tsuchiya et al. [14] investigated the MT and mechanical behavior of TiNi SMA subjected to severe plastic deformation by cold rolling. Their results show the MT suppression beyond the thickness reduction over 25%, which lead to the absence of stress-plateau and small hysteresis in the pseudoelastic stress–strain curves during loading and unloading. Waitz et al. [15] experimentally studied two competing mechanisms in NG TiNi SMAs by TEM, that compensate the transformation strains of the martensite. A single variant of compound twinned martensite forms below a critical grain size of about 100 nm, while a herringbone morphology of two different twinned variants of the martensite is observed in larger grains with additional interfaces causing local strain concentrations. This size dependence of the martensitic morphology is caused by a different scaling behavior of the homogeneous and interfacial strain energies. Ye et al. [16] conducted in situ TEM nanocompression tests to study the stress-induced MT in NiTi nanopillars. The results indicate that stress-induced MT occurs at the stress level about 1 GPa in the pillar sample of diameter below 200 nm. Ahadi and Sun [17] experimentally investigated the effects of grain size on stress hysteresis and temperature dependence of MT stress in NG NiTi SMAs with the average grain size from 10 nm to 1500 nm and found that reducing grain size decreased both stress hysteresis and temperature dependence of MT stress, and finally lead to the disappearance of stress hysteresis and the breakdown of the Clausius-Clapeyron equation for first-order phase transformation. Ahadi and Sun [18] also experimentally investigated the effects of grain size on the rate-dependent thermomechanical responses of NG NiTi SMAs with the average grain size from 10 nm to 90 nm under both monotonic and cyclic tensile loading–unloading. Measurements of stress–strain curves, hysteresis loop area, and temperature fields are synchronized using in situ infrared thermography in the strain rate range from $\dot{\varepsilon} = 4 \times 10^{-5}/s$ to $\dot{\varepsilon} = 0.1/s$. Their experimental results show that in the grain size range, the rate dependence of the transformation stress and the hysteresis loop area gradually weakens and finally tends to vanish for the grain size of 10 nm, and the cyclic stability is greatly improved. Xia et al. [19] reported experimental results on the grain size dependence of Young's modulus and hardness for NG NiTi SMAs with grain size ranging from 10 nm to 120 nm. The hardness of the NiTi SMAs monotonically decreases with the increase in grain size, as predicted by the Hall-Patch equation. Reeve et al. [20] conducted experimental and molecular dynamics simulations to study the MT in $Ni_{63}Al_{37}$ SMA. Their results show that internal stress fields originating from a tailored second phase can change the MT nature from first order to second order phase transformation. The change in MT mode from abrupt to continuous transition is characterized by a corresponding reduction of the thermal hysteresis to zero. In brief, the reported experimental results have confirmed that MT in SMAs depends greatly on the grain size, especially when the grain size is at the nanometer scale.

Along with experimental investigations, theoretical modeling, atomistic calculations, and numerical simulations have been carried out to understand deeper the size-dependent MT in SMAs. For examples, Waitz et al. [21] conducted theoretical study on the size-dependent MT in SMAs, based on two experimental results that 1) the MT proceeds by the formation of atomic-scale twins and 2) grains of a size less than about 50 nm do not transform to martensite even upon large undercooling. In the theoretical study, NGs are modeled as spherical inclusions containing (0 0 1) compound twinned martensite. With the model, they analyzed quantitatively, in terms of energy contributions, the driving force and the resistance to the MT, and hence predicted a critical grain size, about 50 nm for TiNi SMAs, below which the MT becomes unlikely. The theoretical study is also able to predict the morphology of the martensitic phase. Sun and He [22] proposed a one-dimensional (1D) multiscale continuum model to study the grain size dependent MT in NG SMAs. The model indicates that the specific energy dissipation or the width of the stress hysteresis is governed by the ratio of specimen size over the grain size and the ratio of grain size over intrinsic material length. In particular, the stress hysteresis vanishes when the grain size is reduced to the nano-scale where the grain size and the interface thickness become comparable. In bulk specimens of NG SMAs, the specimen size is usually much larger than the grain size and, therefore, only grain size and intrinsic material length will play the considerable role in the size-dependent MT. In addition to theoretical modeling, molecular dynamics (MD) simulations are able to provide physical pictures of MT at the atomic level. The MD simulations [23] show that SMA nanoparticles below the critical size exhibit superelasticity with such features of absence of hysteresis, continuous nonlinear elastic distortion, and high blocking force. The hysteresis-free superelasticity stems from a second-order MT under external stress, which is attributed to a surface effect; i.e., the surface locally retards the formation of martensite and then induces a critical-end-point-like behavior when the system is below the critical size [23]. Besides, surfaces can induce size-dependent Young's modulus [24] and cause surface segregation [25]. The MD simulation results are consistent with the experimental results [26, 27], especially at the point that surface atoms do not participate in the MT. Atoms at interfaces or grain boundaries behave more or less like surface atoms. The scale of atomistic calculations, however, is too small to simulate and understand the experimentally observed microstructure morphology during thermally or/and mechanically induced MT in SMAs and more importantly, the associate thermal/mechanical behaviors. Phase field simulations, based on thermodynamics and kinetics, are able to simulate the evolution of microstructure morphology. Zhu et al. [5, 28-30] conducted a series of phase field simulations, based on the Ginzburg-Landau (GL) theory, of MT in concentration modulated SMAs, exhibiting a high-order phase transition in sharp contrast to the first-order nature of conventional MT and achieving linear-superelasticity and ultralow modulus. The simulations imply that one can artificially manipulated the MT process, extending the repertoire of deformation mechanisms and tailoring mechanical properties of materials. Ahluwalia et al. [31] conducted phase field simulations of the grain size dependent MT and the stress-strain curves of NG SMAs with mean grain sizes of 51.2 nm, 25.6 nm, and 12.8 nm, where the grain boundary thickness (width) was assumed to be 2 nm and the MT was assumed to be an ideal square austenite to rectangle martensite transformation. If the grain boundary energy of the martensite is higher than that of the austenite phase, the MT will not occur in the grain boundary regions. The simulations show that MT in the NG SMAs is suppressed as grain size is decreased, and below a critical grain size, the MT will not occur at all. The stress hysteresis is also greatly reduced with the decrease in grain size. All the simulation results illustrate that the mechanical behavior of NG SMAs is influenced by inter-granular interactions and especially, by grain boundaries. Sun et al. [32] (2018) adopted the square-to-rectangular phase transformation model and considered the inertial effect, the latent heat release and conduction to study the microstructure evolution and the thermomechanical responses of NG SMA nanowires, with grain sizes of 100 nm and 70 nm and grain boundary thickness of 4 nm, under dynamic loadings. The simulation results show that the martensitic volume fraction after the temperature-induced MT in NG SMA nanowires decreases with the reduction in grain size, indicating the constraint effect of grain boundaries on MT. Obviously, the smaller the grain size is, the more distinct of the constraint effect of grain boundaries will be. To study non-isothermal behavior of MT by phase field simulations, Sun et al. [33] used a continuous piecewise function of temperature to replace the widely used single function of temperature in the free energy. The grain boundary thickness was set to be 4 nm and the kinetic coefficient was set to zero in the grain boundaries. Their simulations indicate that the thermomechanical responses of SMAs exhibit distinct grain size effect. The shape memory capacity is lower, the stress hysteresis is narrower, and the cooling capacity is poorer in the NG SMAs with the smaller grain size of 50 nm in comparing the corresponding values in the NG SMAs with the grain size of 100 nm. Sun et al. [34] assumed the grain boundary thickness to be 5 nm to balance the computation burden and the physical reality to study the asymmetric mechanical responses of tetragonal zirconia polycrystals (TZPs), in which the MT means the tetragonal to monoclinic (t-m) phase transformation, under uniaxial tension and compression by phase field simulation with the modified chemical free energy functional based on the fourth-order Landau polynomial [33]. Their results reveal again that the critical stress for the t-m transformation generally increases as the mean grain size decreases. The maximum transformation strain in the stress plateau stage and the residual strain after unloading decrease as the mean grain size decreases. As expected, the tension-compression asymmetry of TZPs is mainly attributed to the dilatational transformation strain accompanying the t-m phase transformation. In brief, the results of the theoretical modeling, atomistic calculations, and numerical simulations have confirmed the experimentally observed size dependent MT. In particular, the phase field simulations imply that the intrinsic material length is the grain boundary thickness.

The significant progress has been achieved in the understanding of the MT behavior in NG SMAs. The MT mechanism in NG SMAs with ultrasmall grain sizes, however, has not been elucidated. The real grain boundary thickness is about one nanometer for normal large angle grain boundaries except of coherent twin boundaries. How does the microstructure of a SMA evolve during temperature- or/and stress-induced MT? Can the thermal and mechanical properties of SMAs be designed and tailored by reducing grain size down to few nanometers? Moreover, the grain size effect of TiNb alloy SMAs remains largely unexplored despite their great potential applications. Especially, TiNb alloys are one of the most promising biomaterials compared to TiNi alloys owing to their low elastic modulus and good biocompatibility [35]. To improve further the mechanical properties of TiNb alloys without losing their good biocompatibility and other advantages, the production of ultrafine-grained TiNb alloys is a proven approach [7, 8, 36]. As a result, a thorough understanding of the MT dependence on the grain size of NG TiNb alloys and the underlying physical mechanisms are essential for the improvement of material properties. Therefore, Ti2448 (Ti–24Nb–4Zr–8Sn–0.1O wt.%), as a typical bioSMA, and phase field simulation, as a powerful numerical approach, are adopted in the present work to investigate the grain size effects on the behaviors of temperature- or/and stress-induced MT, the microstructure evolution, and the thermal and mechanical properties.

## 2. Phase field model

Shape memory alloys Ti2448 undergo reversible MT between austenite and martensite phases during the loading-unloading process or/and the cooling-heating process. To describe the MT, the order parameter $\eta_p$ is employed to characterize the structural phase transition and represents the $p$th martensite variant in the phase field model. The β (BCC, point group m$\bar{3}$m) to α" (orthorhombic, point group mmm) MT in Ti2448 is associated with six crystallographically equivalent deformation modes (or variants) characterized by six stress free transformation strain (SFTS) tensors $\varepsilon_{ij}^{00}(p)$ (or eigenstrains) without considering the internal shuffle [5]. In order to obtain the eigenstrains, the axes of the reference coordinate are chosen as the crystallographic axes of the cubic parent phase. According to the Burgers correspondence [37], the transformation matrices of the six variants are given as:

$$U_1 = \begin{bmatrix} \zeta & 0 & 0 \\ 0 & \frac{\alpha+\gamma}{4} & \frac{\gamma-\alpha}{4} \\ 0 & \frac{\gamma-\alpha}{4} & \frac{\alpha+\gamma}{4} \end{bmatrix}, U_2 = \begin{bmatrix} \zeta & 0 & 0 \\ 0 & \frac{\alpha+\gamma}{4} & \frac{\alpha-\gamma}{4} \\ 0 & \frac{\alpha-\gamma}{4} & \frac{\alpha+\gamma}{4} \end{bmatrix},$$

$$U_3 = \begin{bmatrix} \frac{\alpha+\gamma}{4} & 0 & \frac{\gamma-\alpha}{4} \\ 0 & \zeta & 0 \\ \frac{\gamma-\alpha}{4} & 0 & \frac{\alpha+\gamma}{4} \end{bmatrix}, U_4 = \begin{bmatrix} \frac{\alpha+\gamma}{4} & 0 & \frac{\alpha-\gamma}{4} \\ 0 & \zeta & 0 \\ \frac{\alpha-\gamma}{4} & 0 & \frac{\alpha+\gamma}{4} \end{bmatrix}, \quad (1)$$

$$U_5 = \begin{bmatrix} \frac{\alpha+\gamma}{4} & \frac{\gamma-\alpha}{4} & 0 \\ \frac{\gamma-\alpha}{4} & \frac{\alpha+\gamma}{4} & 0 \\ 0 & 0 & \zeta \end{bmatrix}, U_6 = \begin{bmatrix} \frac{\alpha+\gamma}{4} & \frac{\alpha-\gamma}{4} & 0 \\ \frac{\alpha-\gamma}{4} & \frac{\alpha+\gamma}{4} & 0 \\ 0 & 0 & \zeta \end{bmatrix},$$

where $\alpha = \sqrt{2}b/a_0$, $\zeta = a/a_0$, $\gamma = \sqrt{2}c/a_0$, and $a_0$ is the lattice constant of parent phase while $a, b, c$ are lattice constants along the [100], [010], [001] crystallographic axes of martensite phase respectively. The eigenstrains can be obtained by:

$$\varepsilon_{ij}^{00}(p) = \frac{1}{2}(U_p^T U_p - I), p = 1,2,\ldots,6. \tag{2}$$

Substituting eq. 1 into eq. 2 results in

$$\varepsilon_{ij}^{00}(1) = \begin{bmatrix} \xi^2 - 1 & & \\ & \frac{(\alpha-\gamma)^2 + (\alpha+\gamma)^2}{16} - 1 & \frac{(\gamma-\alpha)(\alpha+\gamma)}{8} \\ & \frac{(\gamma-\alpha)(\alpha+\gamma)}{8} & \frac{(\alpha-\gamma)^2 + (\alpha+\gamma)^2}{16} - 1 \end{bmatrix},$$

$$\varepsilon_{ij}^{00}(2) = \begin{bmatrix} \xi^2 - 1 & & \\ & \frac{(\alpha-\gamma)^2 + (\alpha+\gamma)^2}{16} - 1 & \frac{(\alpha-\gamma)(\alpha+\gamma)}{8} \\ & \frac{(\alpha-\gamma)(\alpha+\gamma)}{8} & \frac{(\alpha-\gamma)^2 + (\alpha+\gamma)^2}{16} - 1 \end{bmatrix},$$

$$\varepsilon_{ij}^{00}(3) = \begin{bmatrix} \frac{(\alpha-\gamma)^2 + (\alpha+\gamma)^2}{16} - 1 & & \frac{(\gamma-\alpha)(\alpha+\gamma)}{8} \\ & \xi^2 - 1 & \\ \frac{(\gamma-\alpha)(\alpha+\gamma)}{8} & & \frac{(\alpha-\gamma)^2 + (\alpha+\gamma)^2}{16} - 1 \end{bmatrix},$$

$$\varepsilon_{ij}^{00}(4) = \begin{bmatrix} \frac{(\alpha-\gamma)^2 + (\alpha+\gamma)^2}{16} - 1 & & \frac{(\alpha-\gamma)(\alpha+\gamma)}{8} \\ & \xi^2 - 1 & \\ \frac{(\alpha-\gamma)(\alpha+\gamma)}{8} & & \frac{(\alpha-\gamma)^2 + (\alpha+\gamma)^2}{16} - 1 \end{bmatrix}, \tag{3}$$

$$\varepsilon_{ij}^{00}(5) = \begin{bmatrix} \frac{(\alpha-\gamma)^2 + (\alpha+\gamma)^2}{16} - 1 & \frac{(\gamma-\alpha)(\alpha+\gamma)}{8} & \\ \frac{(\gamma-\alpha)(\alpha+\gamma)}{8} & \frac{(\alpha-\gamma)^2 + (\alpha+\gamma)^2}{16} - 1 & \\ & & \xi^2 - 1 \end{bmatrix},$$

$$\varepsilon_{ij}^{00}(6) = \begin{bmatrix} \frac{(\alpha-\gamma)^2 + (\alpha+\gamma)^2}{16} - 1 & \frac{(\alpha-\gamma)(\alpha+\gamma)}{8} & \\ \frac{(\alpha-\gamma)(\alpha+\gamma)}{8} & \frac{(\alpha-\gamma)^2 + (\alpha+\gamma)^2}{16} - 1 & \\ & & \xi^2 - 1 \end{bmatrix}.$$

The phase field model of polycrystalline Ti2448 is similar to that for single crystals based on the time-dependent Ginzburg-Landau equation (see Supplementary Information), except that the energy density formulation in the polycrystalline model is dependent on the grain orientation and the grain boundary [32, 38]. To express the total energy of the polycrystalline system, global and local coordinate systems are used to transfer the local eigenstrains and elastic tensor of each individual grain in the above equations into the global one for all grains. The *X-*, *Y-* and *Z*-axes of the global coordinate system are set to be the axes of Cartesian coordinate system, while the axes of the local coordinate systems in each grain are along the crystallographic axes of the cubic phase. The local eigenstrains and elastic tensor can thus be transformed into global coordinates through a rotation tensor $R$:

$$\varepsilon_{ij}^{global00}(p) = R_{ik}R_{jl}\varepsilon_{kl}^{local00}(p) \qquad (4)$$

$$C_{ijkl}^{global} = R_{im}R_{jn}R_{ko}R_{lp}C_{mnop}^{local} \qquad (5)$$

where $R_{ij}$ is the rotation tensor.

For simplicity, a 2D polycrystalline phase field model is constructed in the present work. The 2D plane is chosen as the (001) crystallographic plane of the cubic phase and individual grain is only allowed to rotate around *Z*-axes of the global coordinate system (i.e. [001] direction) by an angle of $\theta$. Thus, $R_{ij}$ is given as:

$$R = \begin{bmatrix} \cos(\theta) & \sin(\theta) & \\ -\sin(\theta) & \cos(\theta) & \\ & & 1 \end{bmatrix}, \qquad (6)$$

In this case, Z-axis related components of eigenstrains do not contribute to XY in-plane components after the in-plane rotation described above. Furthermore, within the plane strain condition, the in-plane eigenstrains of the cubic phase in 2D are modified according to the Poisson effect as follows:

$$\varepsilon_{ij}^{2D00}(p) = \begin{bmatrix} \varepsilon_{11}^{00}(p) + \nu\varepsilon_{33}^{00}(p) & \varepsilon_{12}^{00}(p) \\ \varepsilon_{21}^{00}(p) & \varepsilon_{22}^{00}(p) + \nu\varepsilon_{33}^{00}(p) \end{bmatrix}. \qquad (7)$$

where $\nu$ is the Poisson ratio. After dropping the duplicate 2D eigenstrains, the eigenstrains are expressed as

$$\varepsilon_{ij}^{2D00}(1) = \begin{bmatrix} \xi^2 - 1 + v\left(\frac{(\alpha-\gamma)^2+(\alpha+\gamma)^2}{16} - 1\right) & \\ & (1+v)\left(\frac{(\alpha-\gamma)^2+(\alpha+\gamma)^2}{16} - 1\right) \end{bmatrix},$$

$$\varepsilon_{ij}^{2D00}(2) = \begin{bmatrix} (1+v)\left(\frac{(\alpha-\gamma)^2+(\alpha+\gamma)^2}{16} - 1\right) & \\ & \xi^2 - 1 + v\left(\frac{(\alpha-\gamma)^2+(\alpha+\gamma)^2}{16} - 1\right) \end{bmatrix},$$

$$\varepsilon_{ij}^{2D00}(3)$$
$$= \begin{bmatrix} \frac{(\alpha-\gamma)^2+(\alpha+\gamma)^2}{16} - 1 + v(\xi^2-1) & \frac{(\gamma-\alpha)(\alpha+\gamma)}{8} \\ \frac{(\gamma-\alpha)(\alpha+\gamma)}{8} & \frac{(\alpha-\gamma)^2+(\alpha+\gamma)^2}{16} - 1 + v(\xi^2-1) \end{bmatrix},$$

$$\varepsilon_{ij}^{2D00}(4)$$
$$= \begin{bmatrix} \frac{(\alpha-\gamma)^2+(\alpha+\gamma)^2}{16} - 1 + v(\xi^2-1) & \frac{(\alpha-\gamma)(\alpha+\gamma)}{8} \\ \frac{(\alpha-\gamma)(\alpha+\gamma)}{8} & \frac{(\alpha-\gamma)^2+(\alpha+\gamma)^2}{16} - 1 + v(\xi^2-1) \end{bmatrix},$$

(8)

As a result, the present 2D plane strain model reduces the number of variants from six to four. Based on the experimental results and theoretical analysis [5, 37], the eigenstrains of the above four variants are calculated as

$$\varepsilon_{ij}^{00}(1) = \begin{pmatrix} -0.0545 & 0 \\ 0 & 0.0396 \end{pmatrix}, \varepsilon_{ij}^{00}(2) = \begin{pmatrix} 0.0396 & 0 \\ 0 & -0.0545 \end{pmatrix},$$
$$\varepsilon_{ij}^{00}(3) = \begin{pmatrix} 0.0032 & 0.0260 \\ 0.0260 & 0.0032 \end{pmatrix}, \varepsilon_{ij}^{00}(4) = \begin{pmatrix} 0.0032 & -0.0260 \\ -0.0260 & 0.0032 \end{pmatrix}$$

(9)

Grain boundaries in the polycrystalline model are nontransformable according to experimental observations [39]. Generally, there are two approaches to make grain boundaries nontransformable during the MT of the studied material. The first approach is to constrain the TDGL kinetic coefficient $L$ to be zero at the grain boundaries [40], and the other is to add an external energy to the free energy to suppress the MT at the grain boundaries [31, 32]. In the present work, a grain boundary is regarded as an elastic barrier at which the order parameters do not exist. The value of the order parameters at the interface between the grain interior and grain boundary is assumed to be 0, representing the austenite phase. This treatment is more suitable for the assumption that grain boundaries are nontransformable.

Four polycrystalline models with different average diameters (D~2, 6, 10, 16 nm) and fixed boundary thicknesses of 1 nm are constructed to study the grain size effect on the MT. We randomly select the orientation of individual grain $\theta$ from 0° to 45° with a 5° interval, and other angles outside of this range are equivalent to angles within it due to the crystal symmetry. The geometric topology and the orientation distribution of the polycrystalline models with different average grain sizes are illustrated in Fig. 1(a). The dimensions of the simulation region are set to 64 nm × 64 nm. The zero-flux boundary condition is employed for the order parameters in all the simulations. For the stress-induced MT, the mechanical constraints and uniaxial tensile loading-unloading process are illustrated in Fig. 1(b), in which the loading-unloading stress rate is 100 MPa/s. All the simulation parameters except the eigenstrains are obtained from the literature [5], as listed in Table 1. Based on the selected parameters, the chemical free energy density curves at different temperatures are plotted in Fig. 1(c).

Table 1. Physical meanings and values of the simulation parameters.

| Physical meaning | Value |
| --- | --- |
| Elastic constant | $C_{11} = 57.2$ GPa |
| Elastic constant | $C_{12} = 36.1$ GPa |
| Elastic constant | $C_{44} = 35.9$ GPa |
| Poisson ratio | $\nu = 0.387$ |
| Landau polynomial coefficient | $A = 1.3 \times 10^5$ J/m$^3$ |
| Landau polynomial coefficient | $B = 11.74 \times 10^7$ J/m$^3$ |
| Landau polynomial coefficient | $C = 17.39 \times 10^7$ J/m$^3$ |
| Equilibrium temperature | $T_0 = 133$ K |
| Gradient energy coefficient | $\beta = 1.2 \times 10^{-12}$ J/m |
| Kinetic coefficient | $L = 1 \times 10^{-4}$ m$^3$J$^{-1}$s$^{-1}$ |

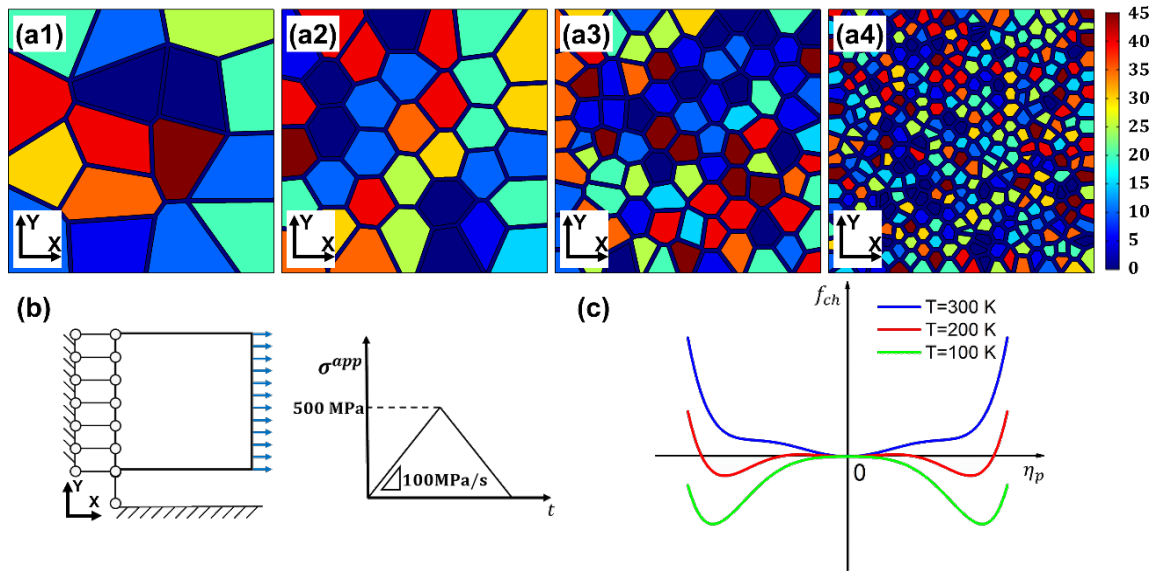

Fig. 1. (a) Topology and orientation distribution of the polycrystalline models with different average grain sizes: (a1) 16 nm; (a2) 10 nm; (a3) 6 nm; (a4) 2 nm. (b) Loading-unloading process of the uniaxial tensile test. (c) Chemical free energy density curves with respect to the order parameter at different temperatures.

## 3. Results and discussion

In this paper, phase field simulations are performed to investigate the grain size effects on the MT behavior using four 2D polycrystalline models with different average grain sizes (D~2, 6, 10, 16 nm). The grain size effects on both the temperature-induced (3.1) and stress-induced (3.2) MTs are studied. Meanwhile, the contribution of the gradient energy to the grain size effect (3.3) and the MT behavior for an ultrasmall grain size (3.4) are further discussed.

### 3.1. Grain size effects on the temperature-induced MT

We first investigate the temperature-induced MT of Ti2448 SMAs with different grain sizes at T= 100 K. Fig. 2 shows the curves of the martensite volume fraction (excluding grain boundary region) versus evolution time for different average grain sizes during the temperature-induced MT. It is clear that the grain size has a significant influence on the suppression of the MT. The maximum martensitic volume fraction of 0.917 for the D~16 nm case is close to the total grain fraction. However, the maximum martensitic volume fraction decreases gradually with decreasing grain size and exhibits a sharp drop for the smallest grain size of D~2 nm. It can also be found that the nucleation time for D~2 nm is longer than that in the other cases, and the D~2 nm case takes more time to complete the MT. Moreover, the martensite volume fraction for D~2 nm gradually increases with time after the start of the MT. These results indicate that a continuous phase transition behavior occurs in the ultrasamll grain size of 2 nm.

To further illustrate the grain size effect on the temperature-induced MT behavior, we compare the microstructure evolutions for D~16 nm and D~2 nm at T=100 K (see Fig. 3). It can not only be seen that the nucleation of D~16 nm martensite is earlier than that of D~2 nm martensite but also that the growth of D~16 nm martensite is faster than that of D~2 nm martensite. Moreover, for the D~2 nm case, the temperature-induced MT occurs successively and individually in different grains, and the martensite in the grain interior tends to be a single martensite variant compared with that for D~16 nm, which is consistent with the experimental results [15]. As shown in the microstructure patterns in the final state (300 ms), the austenite in grains of D~2 nm near the grain boundaries is more obvious than that in grains of D~16 nm. All these results indicate that as the average grain size decreases, the suppression of the MT is strengthened, which corresponds well to the decrease in the maximum martensite volume fraction in Fig. 2.

Generally, the kinetics of the phase transformation consists of different competing mechanisms. For example, the competition between the free energy driving force and the frequency of the atoms with which successfully make the transition from the parent phase to the product phase would produce the "C" shaped isothermal transformation curves (Time-Temperature-Transformation curves, TTT curves) [41]. In our phase field simulation, the kinetics coefficient is assumed to be a temperature independent constant, thus, the frequency of the atoms does not involve in the transition mechanisms. Consequently, the transition kinetics is determined solely by the driving force and the transformation rate $v$ of MT is determined by

$$v = 1/t_x = a(\Delta G(T) - b(D)), \quad (11)$$

where $t_x$ is the time when the transformation fraction reaches $x$, $a$ is the fitting coefficients, $\Delta G(T)$ is the temperature dependent free energy driving force, and $b(D)$ is the grain size dependent coefficient representing the suppression effect of MT. Based on the simulation results (calculated from the simulation results at T=40, 60, 80, 100, 110, 120 and 130 K respectively), the TTT curves with different average grain sizes at which 1%, 20%, and 60% of grain interior area would transform into martensite phase are obtained by fitting eq. (11) and the results are plotted in Fig. 4(a)-(c). The fitting TTT curves are in good agreement with the marked points, which are calculated from the simulation results. Some marked points are missing because the transformation fraction cannot reach that fraction or the temperature-induced MT does not occur at that condition. Due to the single mechanism of the transition kinetics, there is no "C" shaped TTT curves in Fig. 4(a)-(c). The transformation time $t_{0.01}$, $t_{0.2}$ and $t_{0.6}$ increase with the increasing of temperature due to the decreasing of driving force. Meanwhile, it can be also observed that the transformation time increases with the reduction of the average grain size, especially with D~2 nm, which indicates that the driving force is weaken with the decreasing of grain size. Moreover, the fitting coefficient $b$ that represents the suppression of MT, is the biggest in the D~2 nm case and then rapidly decreases with the increase of the average grain size, as shown in Fig. 4(d). The rapid change of the fitting coefficient $b$ shows that the suppression of MT is notable in ultra-small grains, and the grain size effect on the driving force cannot be ignored with ultra-small grain sizes.

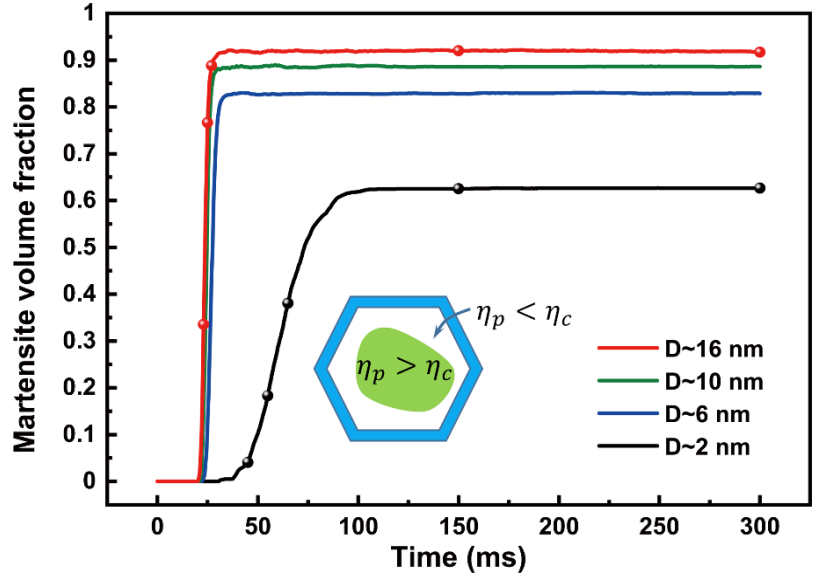

Fig. 2. Martensite volume fraction (excluding grain boundary region) curves during the MT at T=100 K for different average grain sizes. Here, the crystal is regarded as martensite phase when the absolute value of $\eta_p$ is larger than a critical value ($\eta_c$) of 0.4. The microstructure patterns of the marked dots are shown in Fig. 3.

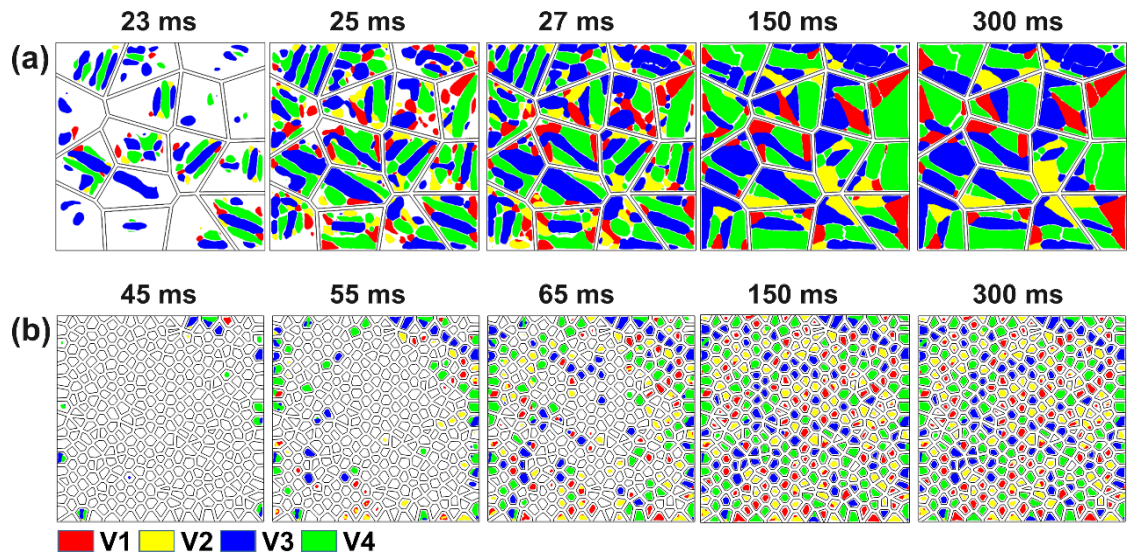

Fig. 3. Microstructure evolution during the temperature-induced MT for an average grain size of (a) D~16 nm and (b) D~2 nm at T=100 K.

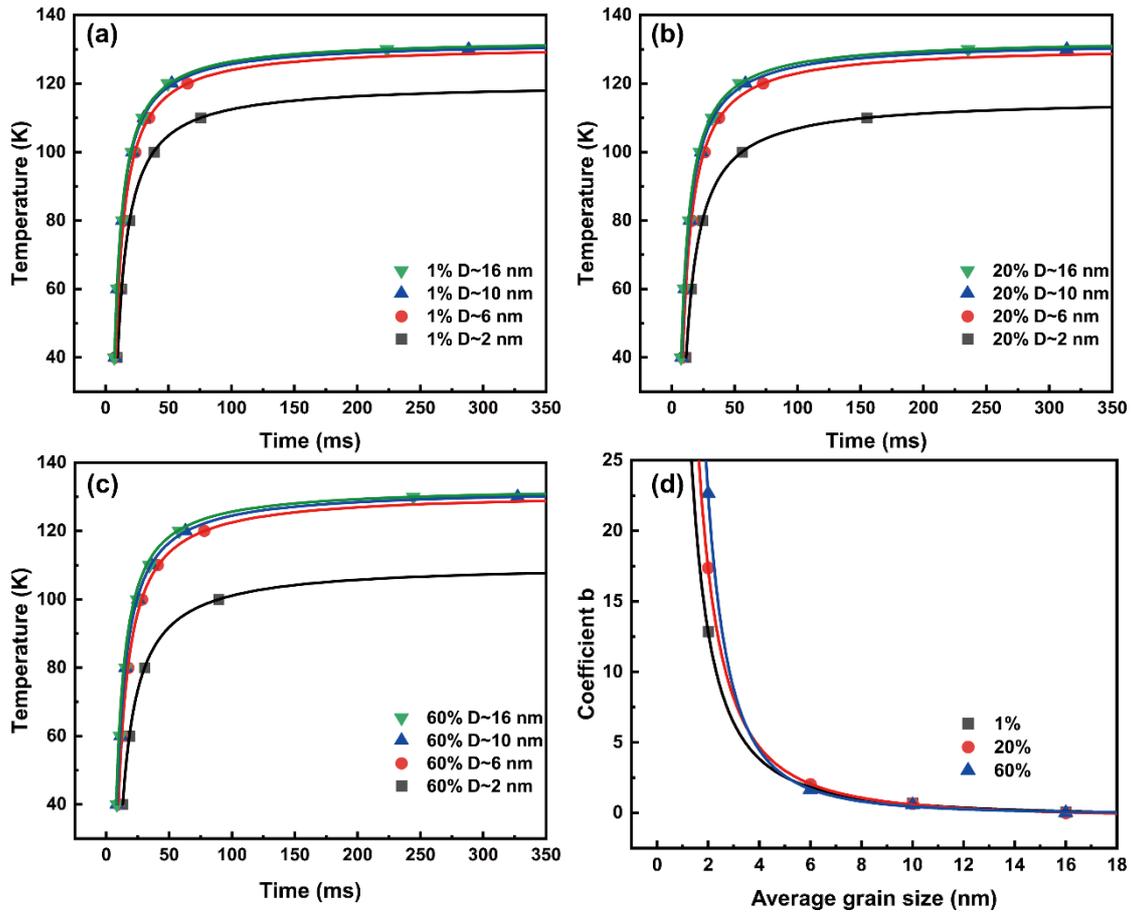

Fig. 4. The isothermal transformation curves (Time-Temperature-Transformation curves, TTT curves) of nano-grained Ti-Nb alloys with different average grain sizes at which (a) 1%, (b) 20%, and (c) 60% of grain interior area would transform into martensite phase. (d) The grain size dependent coefficient $b$ in eq. (11), which represents the strength of the suppression of MT, varies with the average grain size at different transformation fractions.

### 3.2. Grain size effects on the stress-induced MT

#### 3.2.1 Pseudoelasticity

The PE refers to the effect in which an SMA can undergo a large inelastic deformation upon loading while the deformation can be fully recovered upon unloading in a hysteresis loop. During the loading-unloading process, the PE originates from the reversible stress-induced MT when the temperature is higher than the austenite transition temperature. To investigate the grain size effect on the PE, uniaxial tensile tests of the Ti2448 polycrystalline models with different average grain sizes at T=300 K are simulated, and the loading-unloading process is illustrated in Fig. 1(c).

Fig. 5 shows the stress-strain curves under the uniaxial tensile loading-unloading cycle with different average grain sizes at T=300 K. As the average grain size decreases, the critical transformation stress increases, and the pseudoelastic stress hysteresis loop area decreases. Moreover, a typical stress plateau can be observed in all curves except for the D~2 nm case. The obvious difference in the stress-strain curve for D~2 nm compared to the other average grain sizes implies that the stress-induced MT behavior is different for ultrasmall grain sizes.

To clarify the origin of the different mechanical responses of the stress-induced MT behavior, the microstructure evolutions for D~16 nm and D~2 nm under different loading stresses are tracked, and the results are illustrated in Fig. 6, Fig. 7 and Fig. 8. For the D~16 nm case, when the loading stress is above the critical transformation stress (approximately 225 MPa, see Fig. 6(b1)-(b5)), martensite nucleates and grows quickly during the loading process. Even though the MT is inhibited at grain boundaries, the martensite forms a banded pattern across the grain boundaries. Meanwhile, the localized stress concentration around the banded martensite implies that there is an interaction between adjacent grains that can be transferred by the localized stress, as shown in Fig. 7(a). It is worth noting that the grain area cannot be fully filled with martensite at once due to the suppression of the MT arising from the grain boundaries. When the loading stress reaches the maximum loading stress (500 MPa, Fig. 6(b5)), the grains are almost fully filled with martensite, and the corresponding stress concentration occurs near the interfaces between grains and grain boundaries, as shown in Fig. 7(b). Note that only variants 2 and 4 are observed during the loading-unloading cycle because variants 2 and 4 are more favored with orientation angles from 0° to 45° under uniaxial tension along the [100] crystal axis of the parent phase. Since the grain boundaries are nontransformable and assumed to be austenite, the reverse MT avoids the austenite nucleation stage during the unloading process. As the loading stress deceases, the martensite initially shrinks in grains, and then, the martensite disappears in grains successively, as shown in Fig. 6(b6)-(b9). During the unloading process, the reverse MT occurs in grains separately, which is also reflected in the corresponding non-smooth stress-strain curve for D~16 nm (see Fig. 6(a)).

For the D~2 nm case (see Fig. 8), when the loading stress increases to approximately 285 MPa, the MT occurs in one grain, and martensite variant 4 appears. However, this is not obviously reflected in the stress-strain curve because the MT does not occur in the other grains until the loading stress increases to approximately 320 MPa. It is interesting that the orientation of the grain that first starts the MT is not the most favored orientation (0°), indicating that the local environment surrounding the grain has a significant influence on the MT behavior of individual grains, as shown in Fig. 8(a). Meanwhile, it is obvious that the grains start the MT individually under different loading stresses as the loading stress gradually increases, and the grains that start the MT later are not always adjacent to the grains that transformed into martensite earlier. Such continuous phase transition behavior also occurs in the reverse MT during the unloading process. The martensite in grains shrinks and disappears individually as the loading stress decreases, as shown in Fig. 8(b6)-(b9). It is evident that the MT shows a continuous behavior for a small average grain size.

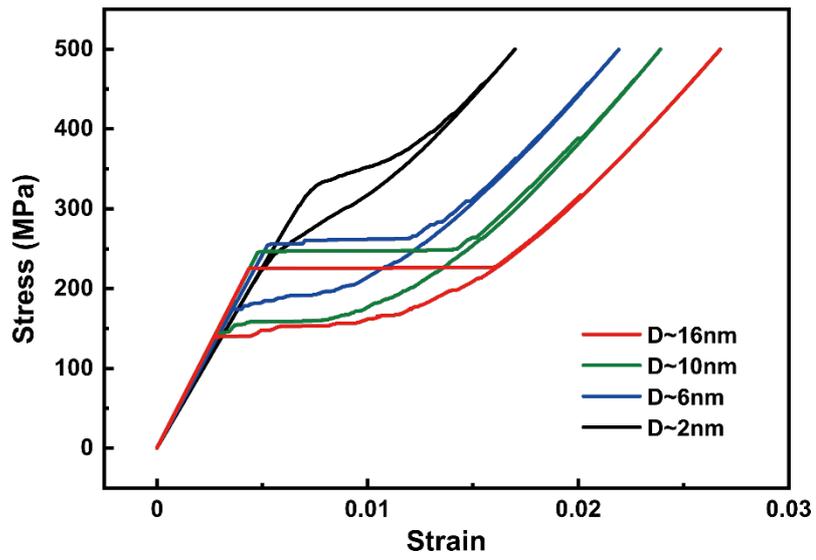

Fig. 5. Strain-stress curves at T=300 K during the loading-unloading process for different average grain sizes.

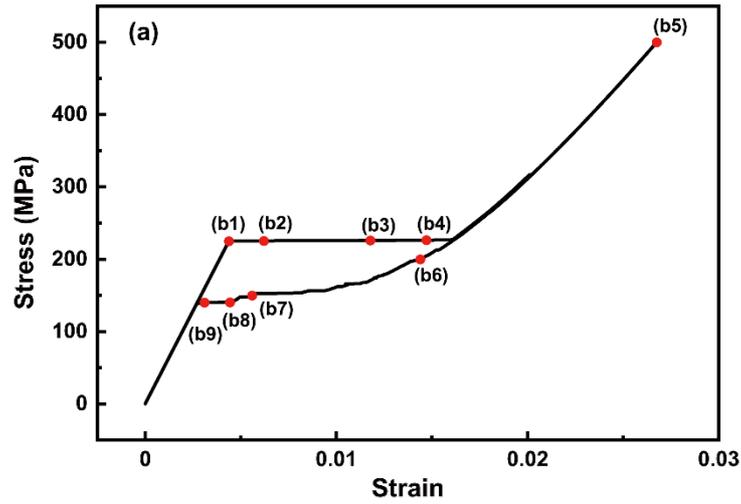

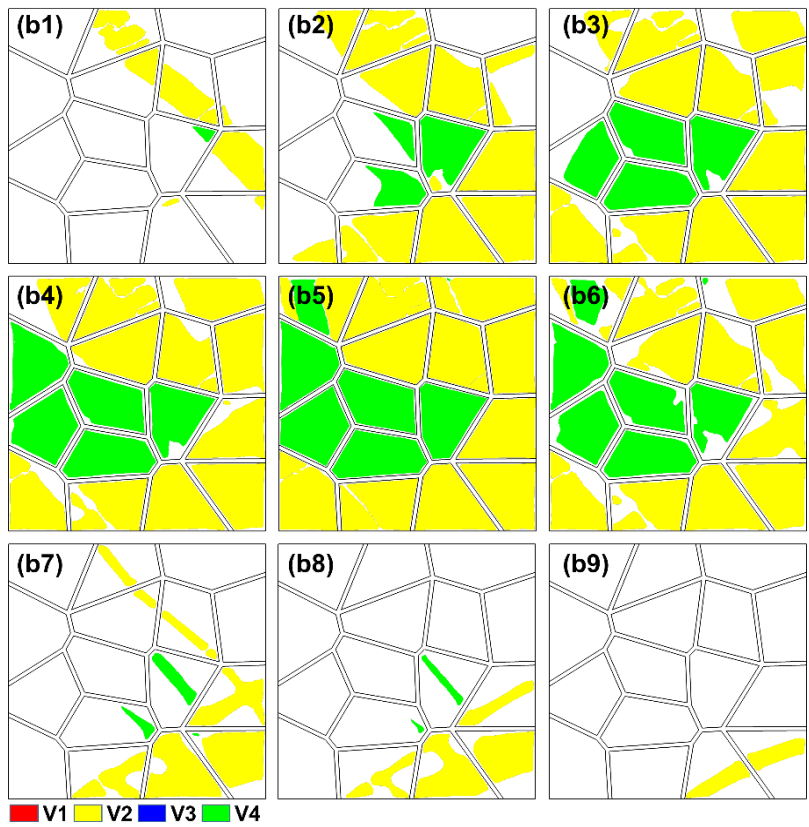

Fig. 6. (a) Stress-strain curve of D~16 nm at T=300 K during the loading-unloading process. The location of (b1)-(b9) on the stress-strain curve is indicated by the red dots. (b) Microstructure evolution patterns for D~16 nm under different loading stresses during the loading-unloading process, loading stage (b1)-(b5): 225 MPa, 225.5 MPa, 226 MPa, 226.5 MPa, and 500 MPa; unloading stage (b6)-(b9): 200 MPa, 150 MPa, 140.5 MPa, and 140 MPa.

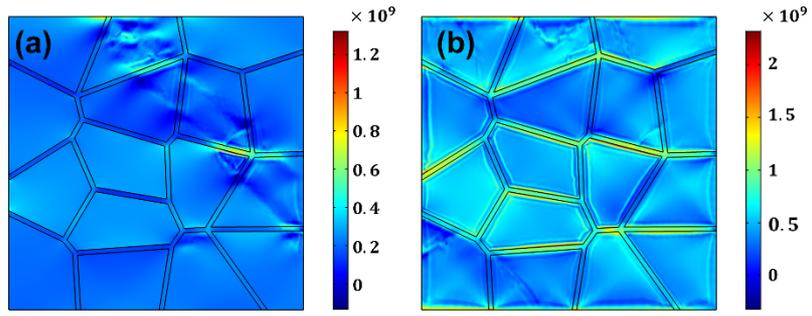
Fig. 7. Contours of $\sigma_{xx}$ (unit: Pa) for D~16 nm in the loading process at (a) 225 MPa and (b) 500 MPa.

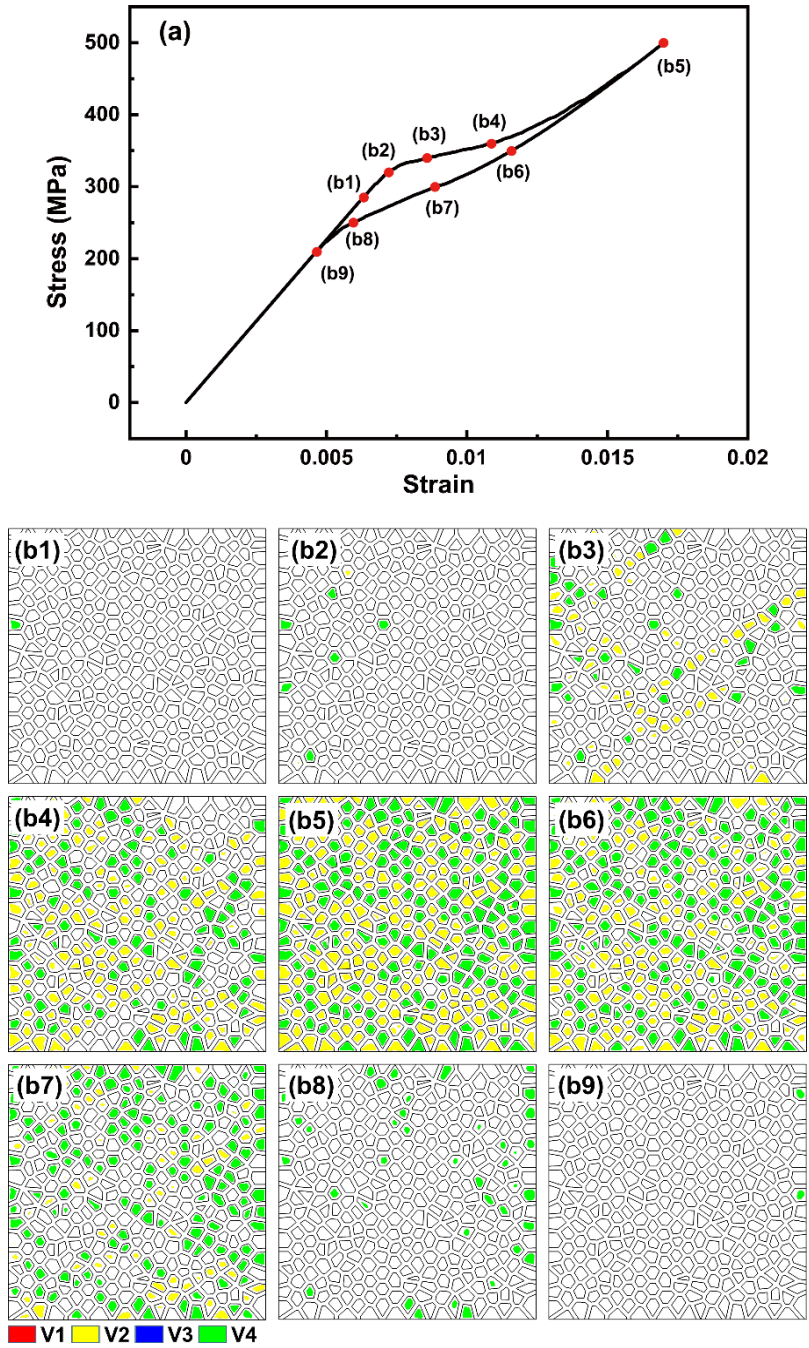

Fig. 8. (a) Stress-strain curve of D~2 nm at T=300 K during the loading-unloading process. The location of (b1)-(b9) on the stress-strain curve is indicated by the red dots. (b) Microstructure evolution patterns for D~2 nm at 300 K under different loading stresses during the loading process (b1)-(b5) and unloading process (b6)-(b9). (b1)-(b5): 285 MPa, 320 MPa, 340 MPa, 360 MPa, and 500 MPa. (b6)-(b9): 350 MPa, 300 MPa, 250 MPa, and 210 MPa.

### 3.2.2 Residual strain

The SME is defined as the recovery of the residual strain that is retrained in SMAs after the loading-unloading process. To investigate the grain size effect on the SME, the residual strains with different average grain sizes are also studied through the cyclic uniaxial tensile loading-unloading process in phase field simulations at T=200 K. The stress-strain curves of nano-grained Ti-Nb alloys upon the first and second uniaxial tensile loading-unloading cycles are shown in Fig. 9(a). As discussed above (in section 3.2.1), the suppression of MT is strengthened in small grain sizes, leading to the increase of the transformation stress and the earlier completion of the reverse MT. Owing to the stronger suppression of MT in smaller grain sizes, the residual strain after the first loading-unloading cycle decreases with the reduction of the average grain size and even disappears in the D~2 nm case, which is also represented in the microstructure patterns shown in Fig. 9(b). These results show that the SME can be modulated at the same working environment temperature by the modulation of grain size. We also found that the stress-strain curves of the second cycle become almost linear without the stress hysteresis area in D~6 nm, 10 nm and 16 nm cases, while the stress-strain curves of the first and second cycles are identical in D~2 nm case. The difference bewteen the stress-strain curves in the second cycle results from that the D~2 nm case does not retain the martensite after the first while the other cases retained (see Fig. 9(b)). The retained martensite could be regarded as the nuclei during the loading process of the second cycle and grows gradually during the loading process, which results in the almost linear and non-hysteresis stress-strain curves for the D~6 nm, 10 nm and 16 nm cases (see Fig. 9(a)). These results thus also provide significant implications for obtaining the linear stress-strain curve by the preexisting martensitic nuclei which can be introduced by pre-strain or modulating the spatial concentration.

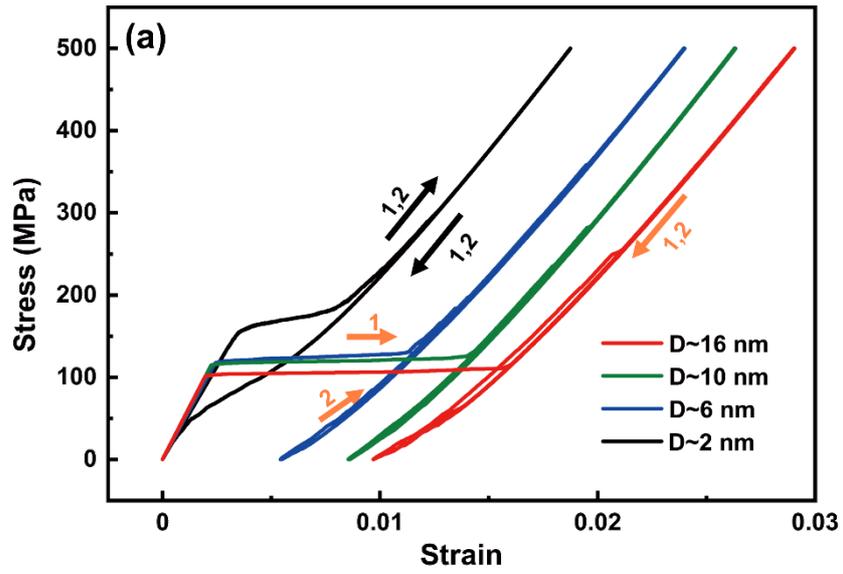

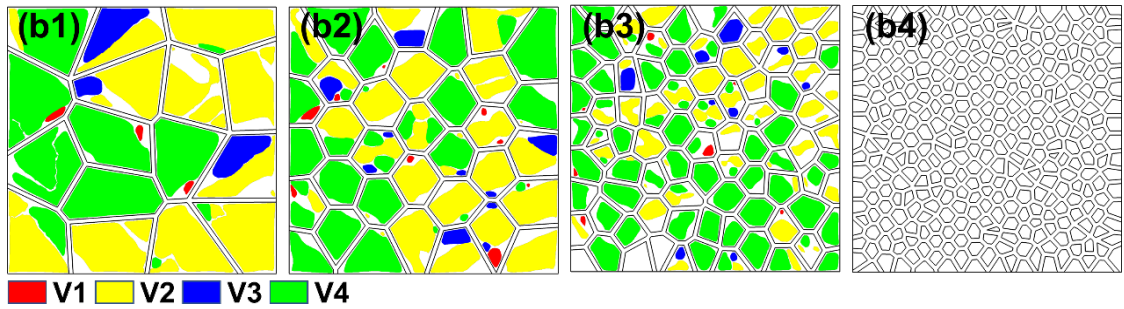

Fig. 9. (a) Stress-strain curves at T=200 K during the first and second cycles of the loading-unloading process for different average grain sizes. The black arrows denote the process of the loading-unloading cycles of D~2 nm, the orange arrows denote the process of the loading-unloading cycles of the other grain sizes. (b) The microstructures after the first loading-unloading process with different average grain sizes. (b1) 16 nm. (b2) 10 nm. (b3) 6 nm. (b4) 2 nm.

### 3.3. Effect of the gradient energy coefficient on the MT

In nanostructured SMAs, the volume fraction of grain boundaries increases greatly as the average grain size decreases. Thus, the gradient energy that arises from the interfaces between the grain boundaries and the grain interiors may dominate the microstructure evolution during the MT for an ultrasmall grain size. Hence, to further study the grain size effect on the MT behavior, we change the gradient energy coefficient directly to investigate its effect on the MT behavior. We perform the phase field simulations with the same conditions and parameters as above except that the gradient energy coefficient is multiplied by a factor of 10 ($\beta = 1.2 \times 10^{-11}$ J/m).

The value of the gradient energy coefficient has an obvious influence on the temperature-induced MT behavior. Compared with the original case (see Fig. 2), a more visible delay of the MT starting moment and a more obvious gap between the martensite volume fraction curves for D~16 nm and D~10 nm can be observed, as shown in Fig. 10(a) for temperature-induced MT at T=100 K with the larger gradient energy coefficient (see Fig. 10(a)). Furthermore, with the larger gradient energy coefficient, the temperature-induced MT is completely suppressed for D~6 nm and D~2 nm, indicating that there exists a critical size, which is also observed experimentally in NiTi [13].

As shown in Fig. 10(b) at T=300 K, with the larger gradient energy coefficient, the grain size effect on the stress-induced MT is more significant. The difference between the critical transformation stresses for different average grain sizes is larger than that in the original case (see Fig. 5). Additionally, the critical transformation stress is much higher than the original stress, indicating enhancement of the suppression of the MT. By comparing Fig. 5 and Fig. 10(b), it can be observed that the stress-strain curve for D~6 nm with a larger gradient energy coefficient is similar to that for D~2 nm with the original parameters, which shows a continuous MT behavior. Moreover, the growth of the martensite in grains of D~6 nm with a larger gradient energy coefficient also shows a continuous phase transition behavior. As shown in Fig. 11, the martensite in the marked grain of D~6 nm nucleates and gradually grows with increasing loading stress. It is clear that as the gradient energy coefficient increases, the grain size effect on the MT behavior is more significant. Therefore, the gradient energy plays an important role in the grain size effect on the MT behavior.

We also note that the nanostructured polycrystalline material with D~2 nm cannot transform into martensite even when the loading stress reaches 500 MPa due to the stronger suppression with the larger gradient energy coefficient. Thus, we increase the maximum loading stress to 1000 MPa and simulate again. As shown in Fig. 12, the stress-strain curve changes slope when the loading stress is near 600 MPa, indicating that the MT has occurred in some grains at that time. The non-hysteresis and the peculiar linearity of the stress-strain curve indicate that the MT and reverse MT occur consecutively. In the zoom-in figure, it is shown that the stress hysteresis becomes very slim. These results imply that the suppression of the MT arising from the gradient energy could further impact the MT behavior.

It has been demonstrated by Hall and Petch in the 1950s [1] that the strength of polycrystalline metals increases with the decreasing grain size according to the following relation:

$$\sigma_Y = \sigma_0 + kD^{-1/2}, \tag{12}$$

where $\sigma_Y$ is the yield stress, $\sigma_0$ is the resistance of the lattice to dislocation motion in grain interior, $k$ is the strengthening coefficient and $D$ is the grain size. This relationship has been most explained by dislocation pile-up mechanism [42]. Our results also show an increase of the transformation stress with grain refinement as discussed above. We then plot the relationship of the inverse square-root of grain size and the transformation stresses extracted from the stress-strain curves in Fig. 5, Fig. 10 and Fig. 12 with different gradient energy coefficients, as illustrated in Fig. 13. The perfect fitting of the above equation indicates that the Hall-Petch law applies to nano-grained Ti-Nb alloys down to the grain size of 2 nm, even though the dislocation effects are not considered. In additional to the well-known dislocation activity, the gradient energy coefficient dependent slope of the fitting relation denotes that the gradient energy is also critical for the Hall–Petch effect. With the decreasing grain size or the increasing gradient energy coefficients, the local stress required to the MT increases due to the confinement of nanograins.

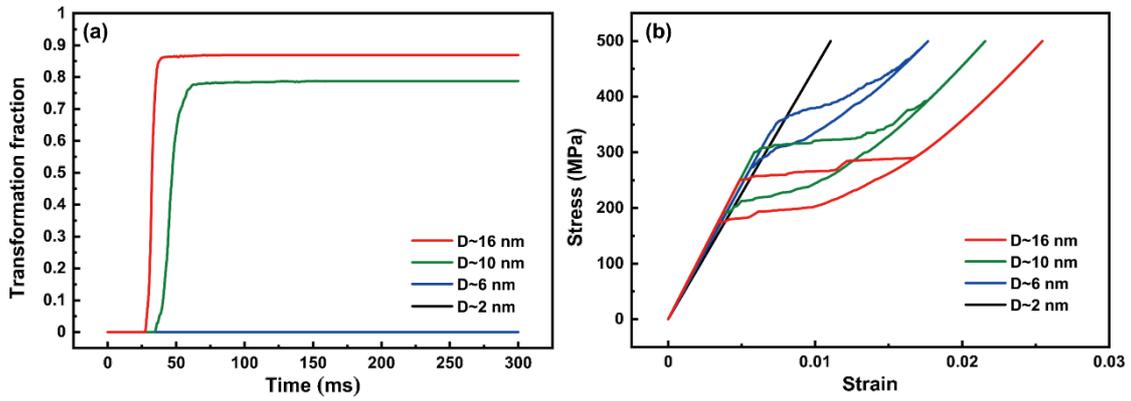

Fig. 10. Simulation results of temperature- and stress- induced MTs with a larger gradient energy coefficient ($\beta = 1.2 \times 10^{-11}$ J/m). (a) Martensite volume fraction (excluding grain boundary region) curves for different average grain sizes during the temperature-induced MT at T=100 K. (b) Stress-strain curves at T=300 K during the uniaxial tensile loading-unloading process for different average grain sizes.

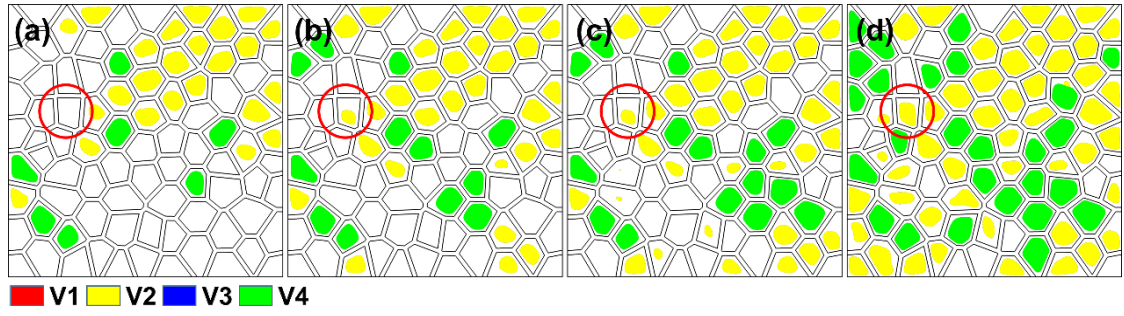

Fig. 11. Microstructure evolution patterns at different loading stresses during the loading process for D~6 nm with the larger gradient energy coefficient ($\beta = 1.2 \times 10^{-11}$ J/m) at T=300 K. (a)-(d): 380 MPa, 390 MPa, 400 MPa, and 450 MPa.

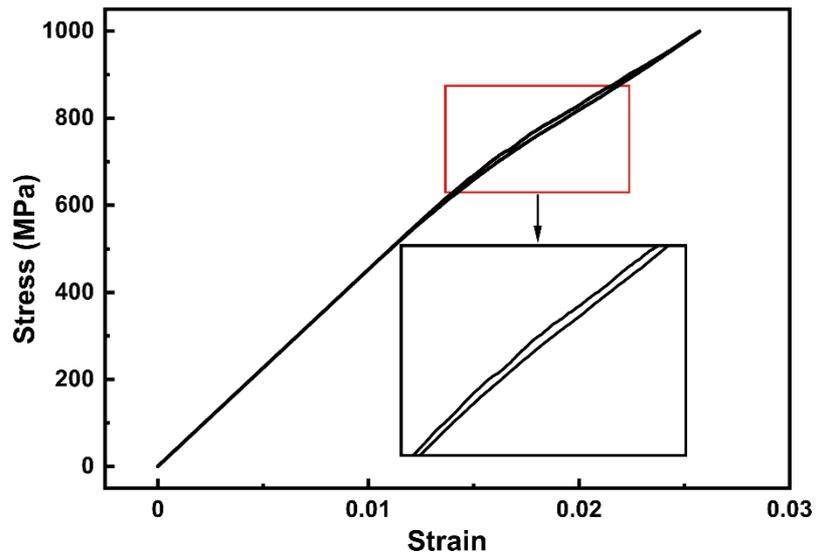

Fig. 12. Stress-strain curves for D~2 nm at T=300 K during the loading-unloading process with the larger gradient energy coefficient ($\beta = 1.2 \times 10^{-11}$ J/m).

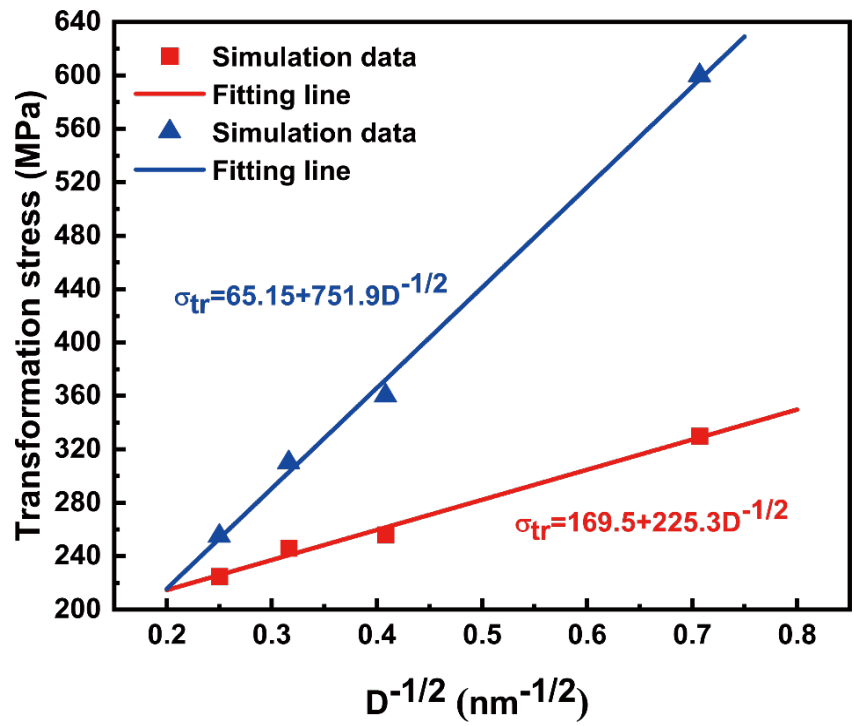

Fig. 13. Plot of the transformation stress at T=300 K against the inverse root of the average grain size ($D^{-1/2}$) for nano-grained Ti-Nb alloys with different gradient energy coefficients.

### 3.4. Discussion

The grain size significantly affects the MT behavior of nanostructured Ti-Nb SMAs with ultrasmall grain sizes. It is illustrated that the stress hysteresis loop and residual strain can be reduced as the average grain size decreases, which benefits the anti-fatigue and engineering design of Ti-Nb medical devices [43, 44]. Moreover, the nanocrystalline Ti-Nb SMAs with ultrasmall average grain sizes or large gradient energy coefficients present a continuous phase transition behavior during the MT. The continuous phase transition behavior manifests in nanocrystalline Ti-Nb SMAs through two different characteristics: one is that grains start the MT individually and not simultaneously, and the other is that the martensite in grains gradually grows with increasing loading stress or evolution time.

The first characteristic results from that each individual grain is situated in a different local environment. Due to the anisotropic mechanical properties, the orientation of individual grains affects the driving force of the MT and the critical transformation stress. Moreover, the geometry of polycrystalline models also impacts the distribution of the local stress concentration. On the other hand, grain boundaries suppress the MT in grains by raising gradient energy. The strength of the suppression of the MT in each grain is different due to the different topology of the grain boundaries around the grains. The difference among the strength of the suppression of different grains is noteworthy with ultrasmall grain sizes or large gradient energy coefficients. Therefore, each individual grain is situated in a different local environment and starts the MT individually in nanostructured Ti-Nb SMAs. The second characteristic results from the dominance of the gradient energy. When the grain size is small or the gradient energy coefficient is large, the gradient energy induced by the interfaces between martensite and austenite is significant. Moreover, the interfaces between martensite and austenite in grains cannot disappear through the MT due to the nontransformable grain boundaries. Therefore, the growth of martensite always increases the gradient energy, which needs to be overcome by the driving force.

According to the above analysis, the suppression of the MT of the grain boundaries, which impacts the MT behavior by increasing the gradient energy, is the major reason for the continuous phase transition behavior. In other words, nanostructured SMAs with large average grain sizes can also present a continuous phase transition behavior when the gradient energy coefficient is sufficiently large. The stress-strain curves with different gradient energy coefficients for D~16 nm shown in Fig. 14(a) are consistent with the analysis. As the gradient energy coefficient increases, the gradient energy becomes dominant for the D~16 nm case, which leads to a continuous phase transition behavior. The two characteristics of the continuous phase transition behavior can be observed in the evolution of the microstructure with a large gradient energy coefficient ($\beta = 1.2 \times 10^{-10}$ J/m), as shown in Fig. 14(b).

The continuous phase transition behavior of the MT can lead to narrow transformation hysteresis and nearly linear PE, which enhances the stability and precise control of SMA devices. Thus, the study of the continuous phase transition behavior of the MT is important for practical applications of SMAs. Recently, a continuous phase transition behavior of the MT was achieved to obtain better properties of Ti-Nb SMAs through concentration modulation [29] and a concentration gradient structure[28]. According to our phase field study, the continuous phase transition behavior can also be achieved by grain refinement, which is meaningful for pursuing excellent Ti-Nb SMA materials.

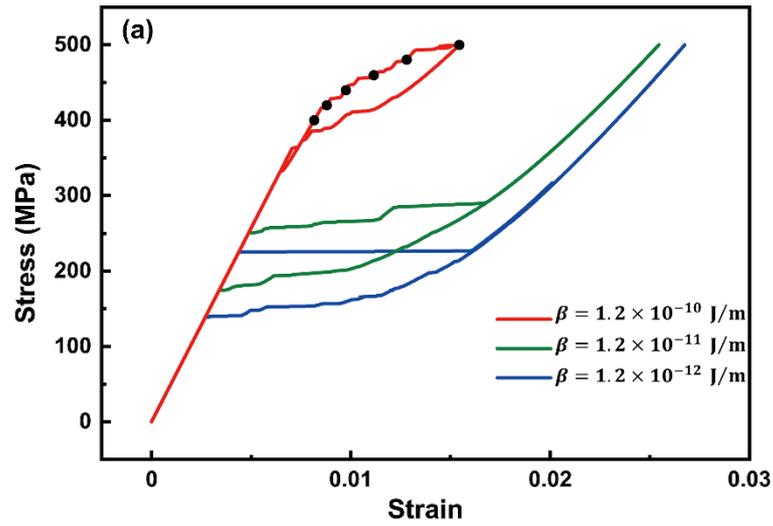
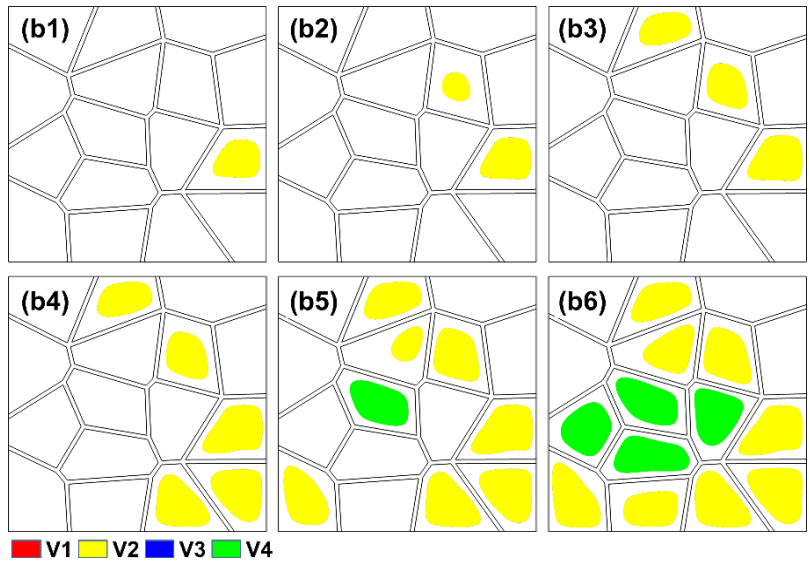

Fig. 14. (a) Stress-strain curves for D~16 nm under different gradient energy coefficients at T=300 K during the loading-unloading process. (b) Microstructure evolution under different loading stresses for D~16 nm with gradient energy coefficient $\beta = 1.2 \times 10^{-10}$ J/m during the loading process. (b1)-(b6): 400 MPa, 420 MPa, 440 MPa, 460 MPa, 480 MPa, and 500 MPa.

## 4. Conclusions

In the present work, phase field simulations are performed to investigate the grain size effect on the MT behavior of nanocrystalline Ti2448. Polycrystalline 2D phase models with different average grain sizes are constructed, in which grain boundaries are regarded as nontransformable elastic barriers. The simulation results exhibit the behaviors of both the temperature- and stress-induced MTs for different average grain sizes and explore the mechanism of the MT behaviors for ultrasmall grain size.

The reduction in the average grain size strengthens the suppression of the MT, which affects the MT behaviors for different average grain sizes. The stress hysteresis and residual strain can be reduced by decreasing the average grain size. The strength of nanocrystalline Ti2448 increases with the decreasing grain size, which agrees well with the Hall-Petch law even the grain size down to 2 nm. With the ultrasmall average grain size, both the temperature-induced MT and stress-induced MT show a continuous phase transition behavior during the MT. The gradient energy plays a critical role in the grain size effect on the MT behaviors. The comprehensive study of the continuous MT transformation behavior in Ti-Nb SMAs with ultrasmall grain sizes will provide theoretical guidance to the manufacture of excellent SMA devices and biomechanical implants.

## Acknowledgements

The authors acknowledge the financial support from the National Natural Science Foundation of China (Grant No. 11802169).